\documentstyle[12pt,subeqnarray]{article}
\begin{document}

\def\tild{\mathop{
\unitlength = 1mm
\begin{picture}(10,3)
\put(2,2){\oval(2,2)[l]}
\put(2,3){\line(3,-1){6.32}}
\put(8,2){\oval(2,2)[r]}
\end{picture}
}
}
\centerline{\large \bf General Approach to
 Functional Forms for }
\centerline{\large \bf the
Exponential Quadratic Operators in}
\centerline{\large \bf Coordinate-Momentum Space  
}

\vskip 1cm

\centerline{Xiang-bin Wang 
\footnote{E-mail address:
scip7236@leonis.nus.edu.sg },
C.H. Oh
\footnote{E-mail address: phyohch@leonis.nus.edu.sg}
and
L.C. Kwek 
\footnote{E-mail address:
scip6051@leonis.nus.edu.sg }
}
\centerline{{\it Department of Physics, Faculty of Science, } }
\centerline{{\it National University of Singapore, Lower Kent Ridge,} }
\centerline{{\it Singapore 119260, Republic of Singapore.  }}

\vskip 0.1in

\vspace{1cm}
\begin{center} \begin{minipage}{120mm}
\begin{center}{ \bf Abstract}  \end{center}
{In a recent paper [Nieto M M 1996 Quantum and Semiclassical Optics,
{\bf 8} 1061; quant-ph/9605032], the one dimensional squeezed and harmonic
oscillator time-displacement operators were reordered in
coordinate-momentum space. In this paper, we give a general approach
for reordering multi-dimensional  exponential quadratic operator(EQO)
in coordinate-momentum space. An explicit computational formula is
provided and applied to the single mode and double-mode EQO through the
squeezed operator and the time displacement operator of the harmonic
oscillator. \\}
\end{minipage}
\end{center}

\newpage
 \section {Introduction}
The  exponential quadratic operator(EQO) plays an important role in
quantum mechanics and quantum optics. In quantum optics, 
such operators occur ubiquitously in topics related to coherent  and 
squeezed states. Consequently, it has always been important to
devise and explore simplifying computational procedures for reducing 
these operators into some manageable forms. In
many applications, 
one usually expresses these operators into their
normal ordered forms. In a recent paper[1], it has been shown that it
is also convenient to consider the reordering of these operators in
coordinate-momentum ($x$-$p$) phase space 
as: $$ \exp[ \delta ] \exp[\alpha
x^{2}]\exp[\beta x \partial ] \exp[\gamma {\partial}^{2} ] $$ where
$\delta ,\alpha  ,\beta$ and $\gamma$ are c-number parameters. Such
reorderings, together with the following identities [1,2]:
\begin{eqnarray}
\exp[c\partial]h(x)=h(x+c)
\end{eqnarray}
\begin{eqnarray}
\exp[\tau x \partial]h(x)=h(xe^{\tau})
\end{eqnarray}
\begin{eqnarray}
\exp[c{\partial}^{2}]h(x)={\frac{1}{[4\pi c]^{1/2}}}
\int\limits_{-\infty}^{\infty}\exp\left[ -{\frac{(y-x)^2}{4c}}\right]h(y)dy,
\end{eqnarray}
facilitate the computations of the
wavefunction. Moreover, as pointed out in ref[1], reordering the
operators in $x$-$p$ space  can be applied to systems[3] with time
dependent potentials such as:
\begin{eqnarray}
V(x,t)=g^{(2)}(t)x^2+g^{(1)}(t)x+g^{(0)}(t)
\end{eqnarray}
Reordering EQO in $x$-$p$ phase space is therefore an interesting problem
that deserves  further investigations.  Following Wei and Norman[4],
Nieto[1] has reduced the one dimensional EQO reordering problem in
$x$-$p$ space into the solution of four coupled first-order differential
equations with four unknowns. However, a direct calculation formula
that relates the EQO to its reordered form is not available.
Furthermore, the results have not been extended to the 
$n$-dimensional case.

In this paper, we start within the framework of $x$-$p$ space and
construct a very general approach which is suitable to reordering
arbitrary mode EQOs to its reordered form in $x$-$p$ space. In the
following section, we will
outline this general approach and summarize the essential
steps. In section 3, we show that this general approach yields 
an explicit
formula for the reordering of arbitrary one-dimensional EQO. The formula
is then applied to the one dimensional squeezed operator 
and time-displacement operator of the harmonic oscillator.  The
results are the same as ref[1], but
unlike ref[1], we need not solve a system of coupled
differential equations. Finally, in section 4,  
we consider the  reordering of
EQO in two dimensions and apply the same technique to the two-dimensional
squeezed  operator and  time displacement operator of the coupled
harmonic oscillator. 

\section {General Approach}
We denote the $n$-dimensional  coordinate and momentum operators as: $$
x=(x_{1}, x_{2} \cdots x_{n}) \hskip 12pt ; \hskip 12pt \partial=ip=(
\partial_{1},  \partial_{2}, \cdots  \partial_n) $$ The commutation
rule for these operators is \begin{eqnarray} [ x_{i}, {\partial}_{j}]=
-\delta_{ij}.
\end{eqnarray}
Without any loss of generality, we
shall consider the following EQO, \begin{eqnarray}U =\exp \left[ \frac
{1}{2} (x,\partial )   \left(
\begin{array}{cc}{D_{1}}&{F}\\{\widetilde{F}}&{D_{2}}
\end{array}           \right)
\left( \begin{array}{c} \widetilde{x} \\ 
\widetilde{\partial} \end{array} \right)\right], \end{eqnarray} 
where  $D_{1}$, $D_{2}$, 
$F$ are $n\times n$ complex matrices and $D_1 = \widetilde{D_1}$ and $D_2 =
\widetilde{D_2}$; the tilde sign denotes the transpose of a matrix.
It is convenient to introduce the symmetric matrix $R$ as $ \left(
\begin{array}{cc}{D_{1}}&{F}\\{\widetilde{F}}&{D_{2}}
\end{array}           \right)$ and operator 
$\hat A \equiv \frac{1}{2} (x,\partial )  R\left( \begin{array}{c} \widetilde{x} \\ 
\widetilde{\partial} \end{array} \right).$
By direct calculations, we first note that if $L$ and $M$
are $n \times n$ complex matrices and $N$ is a symmetric $n
\times n$  complex matrix, then the following identities hold:
\begin{subeqnarray} 
\left[\frac{1}{2} x N \widetilde{x}, \partial M\right]= 
& \frac{1}{2}\left( x N \widetilde{x} \partial  M-\partial  
M x N \widetilde{x}\right)= &-x \cdot NM;  \\
\left[\frac{1}{2} \partial  N \widetilde{\partial}, x M\right]=
& \frac{1}{2} \left(\partial N \widetilde{\partial}x M-xM \partial  N
\widetilde{\partial}\right)= & \partial\cdot NM \label{ident} \\
\left[xL\widetilde{\partial}, xM\right]
& = & x\cdot LM \\ 
\left[xL\widetilde{\partial}, \partial M\right]
& = & -\partial\cdot \widetilde{L}M.
\end{subeqnarray}
From the above identities, one arrives at
\begin{eqnarray}
\left[\hat A, (x,\partial) K\right]=(x,\partial) R\Sigma^{-1}K
\label{impt}
\end{eqnarray}
where $K$ is an arbitrary $2n \times 2n$ complex matrix
and $\Sigma$ denotes $\left(
\begin{array}{cc}{0}&{I}\\{-I}&{0}\end{array}           \right)$ with
$I$ as an $n\times n$ identity matrix.

Using the above formulae and commutation relations, one can recursively
compute the following relations 
\begin{subeqnarray}
 \left[\hat A, (x,\partial) \right] & = & (x,\partial) R\Sigma^{-1}, \\
 \left[\hat A, \left[\hat A, (x,\partial)\right] \right]& = &
(x,\partial) R\Sigma^{-1}\cdot R\Sigma^{-1}
=(x,\partial) \left( R\Sigma^{-1}\right)^2 \\
& \cdots & 
\end{subeqnarray} 
Applying Baker-Campbell-Hausdorff (BCH) relations, one gets 
\begin{eqnarray} U ~ (x, \partial )~   U^{-1}=
(x, \partial) + \left[\hat A,(x, \partial)\right] +\frac{1}{2!}\left[\hat A,\left[\hat A,(x, \partial)\right]\right]+\cdots\end{eqnarray}
which immediately yields
\begin{eqnarray} U ~ (x, \partial )~   U^{-1}= (x, \partial )\cdot \exp\left(R\cdot \Sigma^{-1}\right)\end{eqnarray}
We next denote $T=\left(
\begin{array}{cc}{T_{11}}&{T_{12}}\\{T_{21}}&{T_{22}}
\end{array}           \right) =\exp\left(R\cdot \Sigma^{-1}\right)$ 
then we have
\begin{eqnarray}U ~ (x, \partial )~   U^{-1}= (x, \partial ) \cdot \left(
\begin{array}{cc}{T_{11}}&{T_{12}}\\{T_{21}}&{T_{22}}
\end{array}           \right)   \end{eqnarray}
where $T_{11},T_{12},T_{21}$ and $T_{22}$ are $n\times n$ matrices.
The $T_{ij}, ~ i,j = 1,2$ matrices are not independent. To see this, we
note that the exponential matrix, $\exp \left(R\Sigma^{-1}\right)$, 
satisfies:
\begin{eqnarray}
 \Sigma^{-1}\exp \left(-R\Sigma^{-1}\right)\Sigma =\exp \left(-\Sigma^{-1}R\right)
=\exp \left( \stackrel{\tild}{ ~R\Sigma^{-1}~} \right). \label{prev}
\end{eqnarray}
As $\Sigma^{-1} = - \Sigma$, the above equation (\ref{prev}) becomes
\begin{eqnarray}
\exp \left(\stackrel{\tild}{~R\Sigma^{-1}~} \right) \Sigma \exp \left[\left(R\Sigma^{-1}\right)\right)=\Sigma
\end{eqnarray}
which, in our notation, can be recast as
\begin{eqnarray}
\widetilde T\Sigma T=\Sigma. \label{tmat}
\end{eqnarray}
Expanding and equating the entries in eq(\ref{tmat}), we get
\begin{subeqnarray}
\widetilde {T}_{22}T_{11}-\widetilde{T}_{12}T_{21} & = & 1; \\
\widetilde{T}_{21}T_{11}& = & \widetilde{T}_{11}T_{21}; \\
\widetilde{T}_{22}T_{12} & = & \widetilde{T}_{12}T_{22}. \label{dep}
\end{subeqnarray} 
One can then easily manipulate eq(\ref{dep}) to get the relation
\begin{eqnarray}
T_{11}=\widetilde T_{22}^{-1}+T_{12}T_{22}^{-1}T_{21}
\end{eqnarray}
Furthermore, by these identities, one can always 
have the following decomposition
\begin{eqnarray}  \left( \begin{array}{cc}{T_{11}}&{T_{12}}\\{T_{21}}&{T_{22}}
\end{array}           \right)=  \left( \begin{array}{cc}{I}&{W}\\{0}&{I}
\end{array}           \right) \left( \begin{array}{cc}{e^{Y}}&{0}\\{0}&{e^{-\widetilde{Y}}}
\end{array}           \right) \left( \begin{array}{cc}{I}&{0}\\{Z}&{I}
\end{array}           \right)\end{eqnarray}
with \begin{eqnarray} W=T_{12}{T_{22}}^{-1} ,\;  Z={T_{22}}^{-1}T_{21},
\; Y=-\ln\widetilde{T_{22}}.\end{eqnarray}

 Let $$U_{1} =\exp \left[ \frac {1}{2} (x,\partial )   \left(
\begin{array}{cc}{-W}&{0}\\{0}&{0}
\end{array}           \right)\left( \begin{array}{c} 
\widetilde{x} \\ \widetilde{\partial} \end{array} \right)\right]$$
$$U_{2} =\exp \left[ \frac {1}{2} (x,\partial )   \left(
\begin{array}{cc}{0}&{Y}\\{\widetilde{Y}}&{0}
\end{array}           \right)\left( \begin{array}{c} 
\widetilde{x} \\ \widetilde{\partial} \end{array} \right)\right],$$
and $$U_{3} =\exp \left[ \frac {1}{2} (x,\partial )   \left(
\begin{array}{cc}{0}&{0}\\{0} & {Z}
\end{array}           \right)\left( \begin{array}{c} 
\widetilde{x} \\ \widetilde{\partial} \end{array} \right)   \right].$$
Using eq(11), one has 
\begin{eqnarray*}
U_1(x,\partial )U_1^{-1} & = & (x,\partial )\left( 
\begin{array}{cc}{I}&{W}\\{0}&{I}
\end{array}           \right), \\
U_2(x,\partial )U_2^{-1} & = & (x,\partial ) \left( 
\begin{array}{cc}{e^{Y}}&{0}\\{0}&{e^{-\widetilde{Y}}}
\end{array}           \right), \\
 U_3(x,\partial )U_3^{-1} & =  & (x,\partial )\left( 
\begin{array}{cc}{I}&{0}\\{Z}&{I}
\end{array}           \right)
\end{eqnarray*}
Thus the reordered EQO, $U^{'}=U_{1}U_{2}U_{3}$, satisfies
the following relation \begin{eqnarray} U^{'} ~ (x, \partial ) ~ U^{'-1}= (x,
\partial ) \left( \begin{array}{cc}{I}&{W}\\{0}&{I}
\end{array}           \right) \left( \begin{array}{cc}{e^{Y}}&{0}\\{0}&{e^{-\widetilde{Y}}}
\end{array}           \right) \left( \begin{array}{cc}{I}&{0}\\{Z}&{I} \end{array}           \right)= (x, \partial ) \left(\begin{array}{cc}{T_{11}}&{T_{12}}\\{T_{21}}&{T_{22}}
\end{array}           \right). \label{trans1} \end{eqnarray}
As shown in the Appendix, operator $U^{-1}U^{'}$ commutes with all $x_{i}$ and
$p_{i}$ so that $U$ differs from $U^{'}$ by a c-number factor.  This
factor can be shown to be unity by evaluating the matrix element
between any two states  to $U$ and $U^\prime$ respectively (see
Appendix or ref[5] for details).  Finally, we arrive at the
following formula for reordering the EQOs in an $n$-dimensional $x$-$p$
space: \begin{eqnarray} \exp\left[ \frac{1}{2} (x, \partial ) R \left(
\begin{array}{c} \widetilde{x} \\ \widetilde{\partial} \end{array}
\right) \right]=e^{\frac{1}{2}({\rm tr}
Y)}e^{-\frac{1}{2}xW\widetilde{x}}
e^{xY\widetilde{\partial}}e^{\frac{1}{2}\partial
Z\widetilde{\partial}} \label{trans2}\end{eqnarray} In principle, one
can reorder any  $n$-dimensional EQO in x-p space through
eq(\ref{trans2}).

In summary, one can compute the EQO reordering in $x$-$p$ phase space
according to the following fixed procedure:
\begin{enumerate}
\item Given any EQO, one can rewrite it in the form of eq(6)
to obtain the  matrix $R$, and hence the matrices
$D_{1}$, $D_{2}$ and $F$.
\item One then computes the exponential matrix $\exp(R \cdot {\Sigma}^{-1})=\exp\left( \begin{array}{cc}{F}&{-D_{1}}\\{D_{2}}&{-\widetilde{F}}
\end{array}           \right)$ and obtains the matrix $\left( \begin{array}{cc}{T_{11}}&{T_{12}}\\{T_{21}}&{T_{22}}
\end{array}           \right)$.
\item By eq(19), one can construct $W$, $Z$ and $Y$ explicitly.
\item Finally using eq(21), one arrives at the reordered form.
\end{enumerate}
 
\section {One-Dimensional Application}
We now apply the above results to
one-dimensional problems and the general procedure 
simplifies considerably in this case.
For one dimensional problems, we have
\begin{eqnarray} U=\exp \left[ \frac
{1}{2} (x,\partial )   \left( \begin{array}{cc}{a}&{c}\\{c}&{b}
\end{array}           \right)\left( \begin{array}{c} {x} \\ {\partial} \end{array} \right)\right] 
\end{eqnarray}
where $a$, $b$ and $c$ are all arbitrary c-numbers. Straightforwardly,
we easily get
\begin{eqnarray}\exp(R \cdot {\Sigma}^{-1})=
\exp\left( \begin{array}{cc}{c}&{-a}\\{b}&{-c}
\end{array}           \right)& = &
\left( \begin{array}{cc}{\cosh \theta+c\cdot \sinh\theta/\theta}&
{-a\cdot\sinh\theta/\theta}\\
{b\cdot\sinh\theta/\theta}&{\cosh\theta-c\cdot \sinh\theta/\theta}
\end{array}           \right) \\
&=&
\left( \begin{array}{cc}{T_{11}}&{T_{12}}\\{T_{21}}&{T_{22}}
\end{array}           \right)
\end{eqnarray}
where $\theta =\sqrt{c^{2}-ab}$.
Using eq(19), one obtains
\begin{equation}
W={\frac{-a}{\theta}}\sinh\theta\cdot(\cosh\theta-{\frac{c}{\theta}})^{-1}
;Y=-\ln [\cosh\theta-{\frac{c}{\theta}}\sinh\theta]
;Z={\frac{b}{\theta}}\sinh\theta\cdot(\cosh\theta-{\frac{c}{\theta}})^{-1}.
\label{sub1}
\end{equation}
Substituting eq(\ref{sub1}) into eq(\ref{trans2}) gives
\begin{eqnarray}\begin{array}{l}\exp \left[ \frac
{1}{2} (x,\partial )   \left( \begin{array}{cc}{a}&{c}\\{c}&{b}
\end{array}           \right)\left( \begin{array}{c} {x} \\ 
{\partial} \end{array} \right)\right] 
 ={\frac{1}{\sqrt{\cosh\theta-{\frac{c}{\theta}}}}}\cdot \\ 
\exp[{\frac{1}{2}}{\frac{a}
{\theta}}\sinh\theta(\cosh\theta-{\frac{c}{\theta}})^{-1}x^{2}]
\exp[-\ln(\cosh\theta-{\frac{c}{\theta}})x\partial]
\exp[{\frac{1}{2}}{\frac{b}{\theta}}\sinh\theta
(\cosh\theta-{\frac{c}{\theta}})^{-1}\partial^{2}]
\end{array}
\label{oned1}
\end{eqnarray}

\vspace{6mm}

\noindent Eq(\ref{oned1}) is an explicit formula for reordering any arbitrary
one-dimensional EQO.  

To illustrate the use of eq(\ref{oned1}), 
we consider two specific examples[1]: 
the time displacment operator of the harmonic oscillator and the
squeezed operator in one dimension.  
For time
displacement operator of the harmonic oscillator,
$$T=\exp[\frac{-it}{2}(x^2-{\partial}^2)]$$ Comparing this expression
with eq(\ref{oned1}), we get $$a=-it; \; b=it; \; c=0; \;
\theta=\sqrt{0^2-(-it)\cdot it}=it $$
Using eq(\ref{oned1}), it follows
\begin{eqnarray} T={\frac{1}{\sqrt{\cos t}}}\exp[-{\frac{i}{2}}\tan t
x^2]\exp[-\ln\cos t x\partial]
\exp[{\frac{i}{2}}\tan t {\partial}^2] \label{oned2}, \end{eqnarray}
which is just eq(44) of ref.[1].

The one dimensional squeezed operator is (eq(9) of ref.[1]):
$$S(z)=\exp[-z_{1}(x\partial +1/2)+iz_{2}(x^2+{\partial})/2]$$
which can be rewritten as
 \begin{eqnarray}U =\exp \left[ \frac
{1}{2}
(x,\partial )   \left( \begin{array}{cc}{iz_{2}}&{-z_{1}}\\{-z_{1}}&{iz_{2}}
\end{array}           \right)\left( \begin{array}{c} {x} \\ 
{\partial} \end{array} \right)\right]. \end{eqnarray}
Comparing it with eq(12), 
$$a=b=iz_{2},~ c=-z_{1},~ \theta =\sqrt{{z_{1}}^2+{z_{2}}^2}=r$$
Using eq(\ref{oned1}), one easily sees that
 $$U =\frac{1}{\sqrt{\cosh r +\frac{z_{1}}{r}\sinh r}} 
\cdot \exp\left[  \frac{iz_{2}}{2r}\sinh r 
(\cosh r +\frac{z_{1}}{r}\sinh r)^{-1}x^2\right]\cdot $$
\begin{eqnarray} \exp\left[  -\ln (\cosh r +\frac{z_{1}}{r}\sinh r)x
\partial \right] \exp\left[  \frac{iz_{2}}{2r}\sinh r
(\cosh r +\frac{z_{1}}{r}\sinh r)^{-1}{\partial}^2\right],
 \end{eqnarray}
which is just the eqs.(37) of ref.[1]. 

\section{Two-Dimensional Application}
Finally, we consider the two dimensional problem and reorder some 
two dimensional EQOs in $x$-$p$ space. 
The two-mode squeezed operator is given by[8]
$$S=\exp\left[ g a_{1}a_{2}-g^{*}{a_{1}}^{+}{a_{2}}^{+} \right]$$
Using $({a_{i}}^+,a_{i})=\frac{1}{\sqrt{2}}(x_{i},{\partial}_{i} )   \left( \begin{array}{cc}{1}&{1}\\{-1}&{1}
\end{array}           \right),$ 
we can rewrite this squeezed operator $S$ as
\begin{eqnarray}
S=\exp \left[ \frac
{1}{2}
(x,\partial )  N \left( \begin{array}{cc}(-g^{*})\sigma &0\\0&(g)\sigma
\end{array}           \right)N^{-1}\left( \begin{array}{c} \widetilde{x} \\ \widetilde{\partial} \end{array} \right)\right] \end{eqnarray}
where $\displaystyle \sigma = \left( \begin{array}{cc}
0 & 1 \\
1 & 0 
\end{array} \right),$
$N= \frac{1}{\sqrt{2}}\left( \begin{array}{cc}{I}&{I}\\{-I}&{I}
\end{array}           \right)$ and $I$ is the $2\times 2$ identity matrix. 
Here, both $x$ and $\partial$ are two-dimensional (two modes) vectors. 
Let $R$ be the matrix given by
$$R=N\left( \begin{array}{cc}-g^{*}\sigma &0\\0&g\sigma
\end{array}           \right)N^{-1}.$$ 
It is easy to see that
\begin{eqnarray}
\exp\left( R{\Sigma}^{-1} \right)& = &
N\cdot \exp\left( \begin{array}{cc} 0&g^{*}\sigma\\g\sigma & 0
\end{array}           \right)\cdot N^{-1} \\
& = & \left( \begin{array}{cc} \cosh |g| \cdot I+\frac{g+g^{*}}{2|g|}
\sinh |g| \cdot \sigma & \frac{g^{*}-g}{2|g|}\sinh |g| \cdot \sigma \\ 
\frac{g-g^{*}}{2|g|}\sinh |g| \cdot \sigma&\cosh |g| \cdot I-
\frac{g+g^{*}}{2|g|}\sinh |g| \cdot \sigma  
\end{array}           \right)
\end{eqnarray}
Following our general procedure and denoting
$s_{\pm}$ as $\displaystyle 
\frac{g \pm g^{*}}{2|g|}\sinh |g|$, we get via eq(21)
\begin{eqnarray} \left\{ \begin{array}{l}
W=\frac{-s_-}{{\cosh}^2|g|-s_+^2} \left( 
\begin{array}{cc}{s_+}&{\cosh |g|}\\{\cosh|g|} & {s_+}
\end{array}           \right) \\
Z=\frac{s_-}{{\cosh}^2|g|-s_+^2}\left( 
\begin{array}{cc} {s_+}& {\cosh|g|}\\{\cosh|g|} &{s_+}
\end{array}           \right)\\
Y=-\ln \left( \begin{array}{cc}{\cosh|g|}&{-s_+}\\{-s_+}&{\cosh|g|}
\end{array}           \right) \end{array}\right. \end{eqnarray}
With these quantities solved, one gets from eq(11) the $x$-$p$ reordered
form for the two modes squeezed state operator.

For the time displacement operator of a two-dimensional 
coupled harmonic oscillator with the Hamiltonian
$$H=\frac{1}{2}(x\widetilde{x}+
\partial\widetilde{\partial})+\lambda x_{1}x_{1}, ~ ~ ~ 
-1\leq \lambda \leq 1,$$
we have the time displacement operator 
\begin{eqnarray}U =\exp \left[ \frac
{1}{2}
(x,\partial )  R\left( \begin{array}{c} \widetilde{x} \\ \widetilde{\partial} \end{array} \right)\right] \end{eqnarray}
where $R=\left( \begin{array}{cc}{-itM}&{0}\\{0}&{itI}
\end{array}           \right)$
and  $M= \left( \begin{array}{cc}{1}&{\lambda}\\{\lambda}&{1}\end{array}\right)$.
With these notations, we see that
\begin{eqnarray}
\exp \left(  R {\Sigma}^{-1}\right)= 
\left( \begin{array}{cc}\cos \left( t\sqrt {M}\right)
&i\sqrt{M}\sin \left( t\sqrt {M}\right)\\ {\displaystyle \frac{i\sin 
\left( t\sqrt {M}\right)}{\sqrt{M}}}&\cos \left(t\sqrt {M}\right)
\end{array}           \right)
\end{eqnarray}
where $\sqrt{M}=\left( \begin{array}{cc}{\cos \omega}&{\sin \omega}\\{\sin \omega}&{\cos \omega}
\end{array}           \right)$ and $\omega=\frac{1}{2}\sin^{-1} \lambda$.
Again using eq(9),  we have
\begin{eqnarray}\left\{ \begin{array}{l}
W=   {i}{\sqrt{M}}\tan \left( t\sqrt{M}\right) \\
Z=\frac{i}{\sqrt{M}}\tan \left( t\sqrt{M} \right)\\
Y=-\ln \left[\cos \left( t\sqrt{M}\right) \right] \end{array}\right.\end{eqnarray}
From eq(21),  the $x$-$p$ reordered form 
for the two-dimensional time displacement operator of the 
coupled harmonic oscillator can thus be written down.
  
\section*{Appendix}
In this appendix, we shall show that the operator $U^{-1} U^\prime$
commutes with the position and momentum operators and consequently
$U$ differs from $U^{\prime}$ by a c-number which can be shown to be
unity. We first note that $U$ and $U'$  satisfy the relation
$$U(x,\partial)U^{-1}=U'(x,\partial)U'^{-1}= (x,\partial)T.$$
For the position operator, since
$$U~ x~U^{-1}=U'~x~U'^{-1},$$
we have
$$xU^{-1}U'=U^{-1}U'x.$$
This means $UU'^{-1}$ commutes with all position operators. 
Similarly, one can show that 
$U^{-1}U'$ commutes all momentum operators. 
Clearly, by Schur's lemma, one concludes that $U^{-1} U^\prime$ is
proportional to unity and thus $U'=c\cdot U$.

We next determine the value of $c$. 
Let $|f>$ and $|g>$ be the eigenstate of operator $x$ and $\partial$ 
with zero eigenvalue.
Clearly, $<f|x=0$ and $\partial |g>=0$. Further,  
using the definitition of $U'$, one can immediately see that 
\begin{eqnarray}
<f|U'|g>& = & <f| U_1 U_2 U_3 |g> \nonumber \\
& = & <f| U_2 |g> ~ ~ \mbox{{\rm since}} <f| U_1 = 0 \mbox{\rm ~ and ~} U_3
|g> = 0 \nonumber \\
& = & <f| \exp(-\frac{1}{2} \partial \widetilde{Y} \widetilde{x}) |g>
\nonumber \\
& = & <f| \exp(-\frac{1}{2} \{ x Y \widetilde{\partial} + {\rm tr}~
\widetilde{Y} \} ) |g> \nonumber \\
& = & \exp(-\frac{1}{2}{\rm tr} \ln T_{22})<f|g>. \label{eq34}
\end{eqnarray}
We denote $e^{tR\Sigma^{-1}}$ by the following form: 
$$e^{tR\Sigma^{-1}}=\left( \begin{array}{cc}T_{11}(t)&
T_{12}(t)\\T_{21}(t)&T_{22(t)}\end{array}\right)$$
and proceed to calculate
the  matrix element value of $e^{t\hat A}$ between $<f|$ and $|g>$ as 
$$v(t)=\langle f | e^{t\hat A} |g \rangle. $$ 
However, we note that $\partial |g>=0$ and $<f| x=0$, so that the
derivative $v^{\prime}(t)$ is given by 
\begin{eqnarray}v^{'}(t)=\frac{1}{2} \langle f|( \partial D_2 \widetilde{\partial}
+{\rm tr}F)e^{t\hat A}|g \rangle. \label{deriv}
\end{eqnarray}
In eq(\ref{deriv}), we have used the identity $\partial F \widetilde{x}
= x  \widetilde{F} \widetilde{\partial} + {\rm tr} F$.
From the transformation property of operator $e^{t\hat A}$ in eq(12) we
get the following matrix identity :
\begin{eqnarray}
0 & = & \langle f|e^{t\hat A} \widetilde {\partial} \partial |g
\rangle \\ 
& = & \langle f|[ \widetilde{T}_{12}(t) 
\widetilde{x}+\widetilde{T}_{22}(t)\widetilde{\partial}][xT_{12}t+
\widetilde{\partial} T_{22}(t)]e^{t\hat A} |g \rangle \\
& = & \widetilde{T}_{22}(t)T_{12}(t)v(t)+\widetilde {T}_{22}(t) 
\langle f| \widetilde{\partial} {\partial} e^{t\hat A} |g\rangle 
T_{22}(t) \end{eqnarray}
Without loss of generality, one can assume that 
${\rm det}(T_{22}(t))\not= 0$, so that
      \begin{eqnarray}\langle 0f|  \widetilde{\partial} 
\partial e^{t\hat A} |g\rangle =-T_{12}(t) \label{eq39}
T_{22}(t)^{-1}v(t) \end{eqnarray} 
which leads to
 \begin{eqnarray}
\langle 0|\partial D_2 \widetilde{\partial} e^{t\hat A} |0\rangle 
=-v(t) {\rm tr} [D_{2}
T_{12}T_{22}(t)^{-1}]. \label{eq43} \end{eqnarray}  
Substituting eq(\ref{eq43}) into eq(\ref{deriv})
we get
\begin{eqnarray}v^{'}(t)=v(t) \frac{1}{2} {\rm tr}[\widetilde{F}-D_2 T_{12}
T_{22}(t)^{-1}]. \label{eq44} \end{eqnarray}
On the other hand, one sees that
the derivative of the $T$ matrix is given by
\begin{eqnarray} \frac{d}{dt} \left( \begin{array}{cc} T_{11}(t)&T_{12}(t)\\
T_{21}(t)
& T_{22}(t) \end{array} \right)= \left( 
\begin{array}{cc}F& -D_{1}\\
D_{2}&-\widetilde{F}\end{array}\right)\left( 
\begin{array}{cc} T_{11}(t)&T_{12}(t)\\
T_{21}(t)
& T_{22}(t) \end{array} \right). \label{matreq} \end{eqnarray}
Immediately it follows
\begin{equation}
\frac{d T_{22}(t)}{dt}=D_2
T_{12}(t)-\widetilde{F}T_{22}(t) \label{eq46}
\end{equation}
Putting eq(\ref{eq46}) into eq(\ref{eq44}), one 
sees that $v(t)$ satisfies the
differential equation 
$$v^{'}(t)=-\frac{1}{2}v(t) ~ {\rm tr} [\frac{d T_{22}(t)}{dt}T_{22}(t)^{-1}
]$$
which can be integrated using the condition $v(0)=<f|g>$ to give
\begin{eqnarray}v(t)=\exp[-\frac{1}{2}{\rm tr} \ln T_{22}(t)]<f|g> \label{final}
\end{eqnarray}
Comparing eq(\ref{eq34}) and
eq(\ref{final}) and remembering that $U = {\rm e}^A$, 
the value of $c$-number factor
is unity so that $U=U'$.

\end{document}